\documentclass[12pt]{article}
\pdfoutput=1

\usepackage{bbm}
\usepackage{amsthm}
\usepackage{mathrsfs}
\usepackage{amsfonts, amssymb}
\usepackage{graphicx}
\usepackage{psfrag}
\usepackage[usenames,dvipsnames,svgnames,table]{xcolor}
\usepackage{enumerate}
\usepackage{jcappub}
\usepackage{soul}
\usepackage{subfig}

\newcommand{\beqn}{\begin{eqnarray}}
\newcommand{\eeqn}{\end{eqnarray}}
\newcommand{\beq}{\begin{equation}}
\newcommand{\eeq}{\end{equation}}

\def\simleq{\; \raise0.3ex\hbox{$<$\kern-0.75em
      \raise-1.1ex\hbox{$\sim$}}\; }

\def\simgeq{\; \raise0.3ex\hbox{$>$\kern-0.75em
      \raise-1.1ex\hbox{$\sim$}}\; }

\newif\ifArxivVersion

\ArxivVersionfalse

\title{Inhomogeneous  Anisotropic Cosmology}
\author[a]{Matthew Kleban}
\author[b]{and Leonardo Senatore}
\affiliation[a]{ Center for Cosmology and Particle Physics, Department of Physics, New York University}
\affiliation[b]{ Stanford Institute for Theoretical Physics, Department of Physics, Stanford University and Kavli Institute for Particle Astrophysics and
Cosmology, Department of Physics and SLAC National Accelerator Laboratory}
\emailAdd{kleban@nyu.edu, senatore@stanford.edu}

\begin{document}

\abstract{
In homogeneous and isotropic Friedmann-Robertson-Walker cosmology, the topology of the universe determines its ultimate fate.  If the Weak Energy Condition is satisfied, open and flat universes must expand forever, while closed cosmologies can recollapse to a Big Crunch.   A similar statement holds for homogeneous but anisotropic (Bianchi) universes.  Here, we prove that  \emph{arbitrarily} inhomogeneous and anisotropic cosmologies with ``flat'' (including toroidal) and ``open'' (including compact hyperbolic) spatial topology that are initially expanding must continue to expand forever at least in some region at a rate bounded from below by a positive number, despite the presence of arbitrarily large density fluctuations  and/or the formation of black holes.   Because the set of 3-manifold topologies is countable, a single integer  determines the ultimate fate of the universe, and, in a specific sense, most 3-manifolds are ``flat'' or ``open''.    Our result has important implications for inflation: if there is a positive cosmological constant (or suitable inflationary potential) and initial conditions for the inflaton, cosmologies with ``flat'' or ``open''  topology must expand forever in some region at least as fast as de Sitter space, and are therefore very likely to begin inflationary expansion eventually, regardless of the scale of the inflationary energy or the spectrum and amplitude of initial inhomogeneities and gravitational waves.  Our result is also significant for numerical general relativity, which often makes use of periodic (toroidal) boundary conditions.}


\maketitle

\section{Introduction}

Nearly all work in cosmology focuses on the regime in which the universe is approximately homogeneous and isotropic, plus small fluctuations that can be treated perturbatively.  This is an accurate description to the universe on large scales today.  However, the universe on smaller scales has non-peturbatively large density fluctuations, and the very early universe prior to inflation was presumably highly anisotropic and inhomogeneous.   Indeed, inflation is invoked to explain today's large-scale homogeneity~\cite{Guth:1980zm,Linde:1981mu,Albrecht:1982wi,Linde:1983gd}.  Because inflation is quasi-de Sitter spacetime and de Sitter spacetime is homogeneous, it seems that inflation might require homogeneity to begin.  If so, it may not explain homogeneity at all.

Very recently, we participated in numerically evolving the full non-linear 3+1D Einstein equations for an inflaton  field coupled to gravity, with the initial conditions being a compact universe with toroidal topology and large inhomogeneities \cite{East:2015ggf}.  Due to numerical limitations we could only include inhomogeneities with wavelength $\mathcal{O}(10)$ times the initial Hubble radius and smaller.  These simulations showed for the first time that the inhomogeneities clump into overdense regions that collapse and form black holes, while the rest of the universe continues to expand and eventually begins inflation.  These results indicate that large perturbations on length scales smaller than the initial Hubble radius and up to 10 times larger do not prevent inflation from beginning.

However, we were not able to treat modes with wavelength much larger than the initial horizon size.  These are potentially a big problem, because  even if the energy scale of inflation is as large as the data allows, there is a ratio of at least $10^5$ between a Planck length initial horizon size and the would-be inflationary horizon size.  This means that a very large number of modes will enter the horizon before inflation could possibly begin (at least in a large or non-compact universe where such modes exist).  This certainly sounds worrying.  Suppose one considers a region that is initially expanding.  As the horizon grows, new modes enter.  Each time a new mode enters, one might expect it has a 50\% chance of overclosing that region, causing it to recollapse in about the current Hubble time.  This  suggests that any given region is very unlikely to grow large enough for inflation to begin -- and the lower the scale of inflation, the more modes enter and smaller this probability becomes (this is the ``initial conditions problem" described in \emph{e.g.}~\cite{Ijjas:2015hcc}).

In \cite{East:2015ggf}, we sketched a heuristic argument that indicates that, for an initially expanding universe but even in the presence of large-amplitude superhorizon modes, given one global condition there should always  be at least one region that expands enough to begin inflation.
The argument goes as follows.  Consider a large expanding region -- as large as the would-be inflationary Hubble length, or as large as the universe if it is finite and smaller than this size -- that is ``flat on average" or ``open on average".  Heuristically, this means that the  expansion rate averaged over that region is equal to or exceeds the average density, in the appropriate units.\footnote{Somewhat more formally, it means that the averaged region is described by a Bianchi universe that is not type IX.}  This requirement of flat or open is one global condition that might be expected to hold true perhaps $\sim 50\%$ of the time in a randomly chosen universe or large region within a universe.\footnote{In fact as we will see, in a sense nearly all cosmologies satisfy it.}   Within this large region there may be sub-regions that are closed and collapse before inflation can begin.  However, in any subdivision of the original flat or open volume there must be at least one subregion that is again flat or open on average.  Subdividing  that region, again there must be at least one open or flat sub-subregion, etc.~on down to the initial Hubble length.  Hence, there is at least one initial Hubble volume that will not be crunched by a mode entering the horizon before inflation can begin.

In this paper, we  prove a rigorous
version of that statement.

\subsection{Relation to previous work}

In \cite{Wald:1983ky}, Wald proved that initially expanding homogeneous but anisotropic (Bianchi) cosmologies with $\Lambda > 0$ must asymptote to de Sitter spacetime, with the exception of Bianchi type IX -- which are ``closed'' in the  general sense we define below.  In particular, only Bianchi Type IX can recollapse.  As such, this work can be regarded as a major generalization of \cite{Wald:1983ky} to the case of inhomogeneous cosmologies.  A useful review of maximal surfaces in general relativity can be found in \cite{bartnik1987}.

\smallskip
\paragraph*{\bf Note added:}
After this work was completed and  the paper submitted to the arxiv, the existence of \cite{barrow1985closed} was brought to our attention, of which we were previously unaware.  This work has very substantial overlap with ours: it shows, using  similar techniques, that there exists no maximal hypersurface for universes with compact spatial topologies except the ones we refer to as type (i) (page 3), and  therefore such cosmologies have no global crunch. However, there is no explicit discussion in \cite{barrow1985closed} of the growth of volume of  hypersurfaces (Eq.~(\ref{volgrowth})), or  that there is always a region with expansion rate bounded from below (Eqs.~(\ref{krho}, \ref{ds})), and of the resulting consequences for inflation.

\section{A ``no big crunch'' theorem}

Our result requires three assumptions:

\begin{itemize}

\item A  ``cosmology'', defined to mean a connected 3+1 dimensional spacetime with a compact Cauchy surface (we comment on the non-compact case in Sec.~\ref{noncompact}).  This implies the spacetime is topologically  $R \times M$ where $M$ is a 3-manifold, and that it can be foliated by a family of topologically identical Cauchy surfaces $M_t$ \cite{Geroch}.

\item The  topology of the spatial 3-manifolds must not be ``closed,'' meaning the topology of  $M$ must \emph{not} be of  type (i) in the classification we introduce in Sec.~\ref{proof} (roughly, $M$ cannot be a spherical space).

\item The Weak Energy Condition (WEC) must be satisfied everywhere: $T_{\mu \nu}k^{\mu}k^{\nu} \geq 0$ for all time-like vectors $k^{\mu}$.  This condition is satisfied for matter, radiation, positive cosmological constant, and scalar fields with potential energy $V(\phi) \geq 0$, but not by negative $V$  or negative cosmological constant.

\end{itemize}

With these assumptions we will prove the following statements:
\bigskip

\noindent {   \emph{There cannot exist a non-singular spacelike hypersurface with maximum volume: given any time slice, there is another with larger spatial volume.  Furthermore, in an initially expanding universe there must be at least one expanding region on every timeslice, and if $\Lambda > 0$ the expansion rate in  that region is bounded from below by that of de Sitter spacetime in the flat slicing.} } \smallskip
\bigskip

 This implies that in a big bang cosmology, there cannot be a big crunch (a global collapse to zero volume).
In the context of inflationary cosmology, this strongly suggests that inflation will start no matter how large are the initial inhomogeneities or how small is the energy scale of inflation, at least so long as the inflaton takes a suitable value in the inflationary part of the potential in the regions that expand (see~\cite{East:2015ggf} for more on this last point).

The proof (explained in more detail in the next section) is very simple.  A hypersurface of maximum volume has vanishing traced extrinsic curvature $K=0$.  Certain 3-manifold topologies require that the spatial curvature $R^{(3)} < 0$ somewhere on the manifold.  We will show that the combination of these two facts is incompatible with Einstein's equations and the WEC ($T_{\mu \nu}k^{\mu}k^{\nu} \geq 0$ for timelike $k$).  Therefore, no maximum volume hypersurface can exist.  A minor elaboration on this argument establishes the lower bound for $\Lambda > 0$.

To establish notation, a timeslice $M_{t}$ has induced metric $h_{\mu\nu}=g_{\mu\nu}+n_\mu n_\nu$, where $g_{\mu\nu}$ is the spacetime metric and $n_\mu$ is  orthonormal to $M_t$, $n_{\mu}n^{\mu} = -1$. The extrinsic curvature is $K_{\mu\nu}$, satisfying $n^{\mu} K_{\mu \nu} = 0$ and with trace $K= h^{\mu\nu}K_{\mu\nu} = g^{\mu \nu} K_{\mu \nu}$, and traceless part $\sigma_{\mu \nu} \equiv K_{\mu \nu} - {1 \over 3} K h_{\mu \nu}$.  The three-dimensional Ricci curvature is~$R^{(3)}_{ij}$, with trace $R^{(3)}$.  Our sign convention is mostly plus, so that  $K>0$ corresponds to expansion.  If there is a cosmological constant (which must satisfy $\Lambda \geq 0$ for the WEC), we include it in $T_{\mu \nu}$.

\subsection{Proof of the theorem}\label{proof}
The proof follows almost immediately from a form of the Hamiltonian constraint that can be derived from the Gauss-Codazzi relation (see for instance \cite{Wald84}, eq.~(E.2.27)):\footnote{For
reference, note that in the FRW case \eqref{constraint} becomes (with all terms in the same order)
\begin{equation*}
 16 \pi G_{N} \rho = 6 {k \over a^{2}} + {2 \over 3} 9 H^{2} - 0 ,
\end{equation*}
where $\rho$ is the energy density, $a$ is the scale factor, $H=\dot a/a$, and $k=\pm1,0$ as usual.   If $\rho > 0, H^{2}$ cannot vanish unless $k=+1$.}
\beq \label{constraint}
16\pi G_N  T_{\mu \nu}n^{\mu}n^{\nu} =  R^{(3)} + {2 \over 3} K^2 - \sigma_{\mu \nu} \sigma^{\mu \nu} .
\eeq

 If some hypersurface $M_{t}$ is extremal -- that is, if its volume form $\sqrt{h}$ is  stationary under  local variations of the surface -- then $K=0$ everywhere on it. This holds because
  \beq \label{normdeform}
{\cal{L}}_{n} \log\sqrt{h}=K,
 \eeq
where ${\cal{L}}_{n}$ is the Lie derivative along the normal to the surface.
If  there is a hypersurface with $K=0$ everywhere and $R^{(3)} < 0$ at any point,  the right-hand side of \eqref{constraint} is strictly negative at that point, violating the weak energy condition.\footnote{Note that $\sigma_{\mu \nu} \sigma^{\mu \nu}  =  \sigma_{\mu \nu} \sigma_{\rho \lambda} g^{\mu \rho} g^{\nu \lambda} = \sigma_{\mu \nu} \sigma_{\rho \lambda} h^{\mu \rho} h^{\nu \lambda} \geq 0$, because $h_{\mu \nu}$ is spacelike.}  As we  explain below, for ``most'' 3-manifold topologies $R^{(3)}$ {\it must} be  negative somewhere.  Therefore, if the WEC holds, no time-slice with extremal volume can exist in a universe with those spatial topologies.

\begin{figure}
\begin{center}
\includegraphics[width=14cm,draft=false]{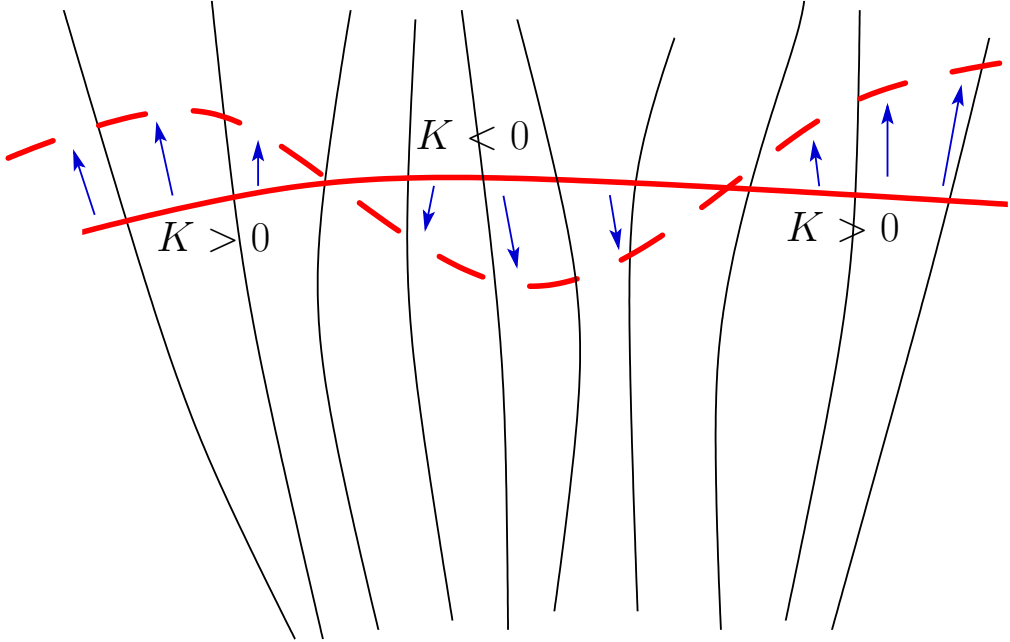}
\end{center}
\caption{
Deforming towards an extremal surface with $K=0$. \label{fig:geometric-construction} }
\end{figure}

In a big bang/big crunch spacetime that expands from zero volume and then globally recollapses, any given foliation must contain a  hypersurface $M_{t}$  that has maximum total volume  $\int_{M_t} \sqrt h$.  However there is no guarantee that $M_t$ is extremal -- that is, it may have regions with $K>0$ and others with $K<0$.   To arrive at a timeslice with $K=0$, consider infinitesimally deforming some  given slice at each point along its normal direction by an amount  $K$.\footnote{The deformation procedure described here corresponds to a procedure known as ``mean curvature flow'', which has been much studied in the mathematics literature for Riemannian manifolds.}  From \eqref{normdeform}, this  guarantees that $\sqrt{h}$  increases pointwise by an amount proportional to $K^{2}$.  Therefore  the total spatial volume
$V \equiv \int d^3 x \sqrt{h}$ satisfies
\beq \label{volgrowth}
{\partial V \over \partial \lambda} = \int d^3 x K^2 \sqrt{h} \equiv \langle K^2 \rangle \geq 0
\eeq
where $\lambda$ is the affine parameter of the deformation.  Hence after the deformation the new surface has either strictly larger volume, or equal volume if and only if $K=0$ everywhere and the initial surface was already extremal.  Iterating this procedure then leads to two possible outcomes:  either it eventually produces an extremal surface with $K=0$ everywhere for which the total volume is maximized, or the volume of the surface continues to grow.
  See Fig.~\ref{fig:geometric-construction} for a pictorial representation of this procedure.\footnote{Formal proofs of the existence of a $K=0$ slice under  the assumption there is  a big bang and a big crunch can be found in~\cite{gerhardt1983} and~\cite{ecker1991}. }

To recap, so far we have established that either a hypersurface with $K=0$ exists, or the volume of space grows  either towards the future or the past, so that an initially expanding spacetime cannot everywhere recollapse.  If a surface with $K=0$ exists, $R^{(3)}$ cannot be negative anywhere without violating the WEC.  To determine under what circumstances  $R^{(3)}$ must be negative somewhere, note that all compact, oriented 3-manifolds fall into one of three topological classes (see \cite{besse1987einstein} Theorem 4.35 and \cite{thurston1997three}):
\begin{itemize}
\item [(i)] ``Closed'':  Manifolds where the scalar curvature is entirely unconstrained  -- any smooth function is the scalar curvature of some smooth metric on the manifold -- and in particular, the scalar curvature can be positive everywhere.  This includes $S^3$, $RP^3$,  and $S^{2} \times S^{1}$, and  more generally $S^{3} / \Gamma$ where $\Gamma \subset SO(4)$ is finite and acts freely (with no fixed points) on $S^{3}$, and connected sums of  such manifolds.
 For these topologies $R^{(3)}$ can be  positive everywhere and our proof does not apply.

\item [(ii)] ``Flat'': Manifolds for which a smooth function is the scalar curvature of some metric if and only if it is either zero identically and the manifold is Ricci-flat, or is negative somewhere.  This includes manifolds of the form $\mathbb{R}^{3}/\Gamma$, where $\Gamma$ is a freely acting finite subgroup of the isometry group of flat Euclidean 3-space (such as torii $T^{(3)}$).

\item [(iii)] ``Open'':  Manifolds for which a smooth function is the scalar curvature of some metric if and only if it is negative somewhere.  This includes manifolds of the form $H^{(3)}/\Gamma$, where $\Gamma$ is a freely-acting subground of the isometry group of $H^{(3)}$ with the standard metric, as well as  $H^{(2)} \times R$, nil, solv, and $\tilde{SL}(2,R)$ manifolds, and connected sums of arbitrary members of all three of the classes above that contain at least one factor of the ``open'' type.

\end{itemize}

As is apparent from this classification the set of compact, orientable three-manifold topologies is countably infinite, meaning (under only the assumption of global hyperbolicity) the topology of the universe can be specified by a single integer $k$ -- just as in the homogeneous and isotropic case, albeit with infinitely many possibilities rather than just $\pm 1$ or $0$.   Furthermore, while there are infinitely many manifolds of type (i), in a sense most 3-manifolds are of type (iii) -- because all connected sums including even one  factor of type (iii) are of type (iii).

For manifolds of type (iii), there must be a region where $R^{(3)} < 0$,  establishing the theorem.  The same is true for universes of type (ii) except in the trivial case of an empty universe, because either the curvature is negative somewhere -- as in type (iii) -- or it is identically zero.  If $R^{(3)} = K = 0$, the only possibility consistent with the WEC is $\sigma_{\mu \nu} = T_{\mu\nu}n^{\mu} n^\nu = 0$ everywhere in the universe, which -- assuming now the dominant energy condition -- implies that all components of $T_{\mu \nu}$ vanish.

Note that in all cases, in any region where $R^{(3)} \leq 0$,

\beq \label{krho}
|K|\geq K_{\star} \equiv \sqrt{24\pi G_N T_{\mu\nu}n^\mu n^\nu}.
\eeq

\begin{figure}
\begin{center}
\includegraphics[width=14cm,draft=false]{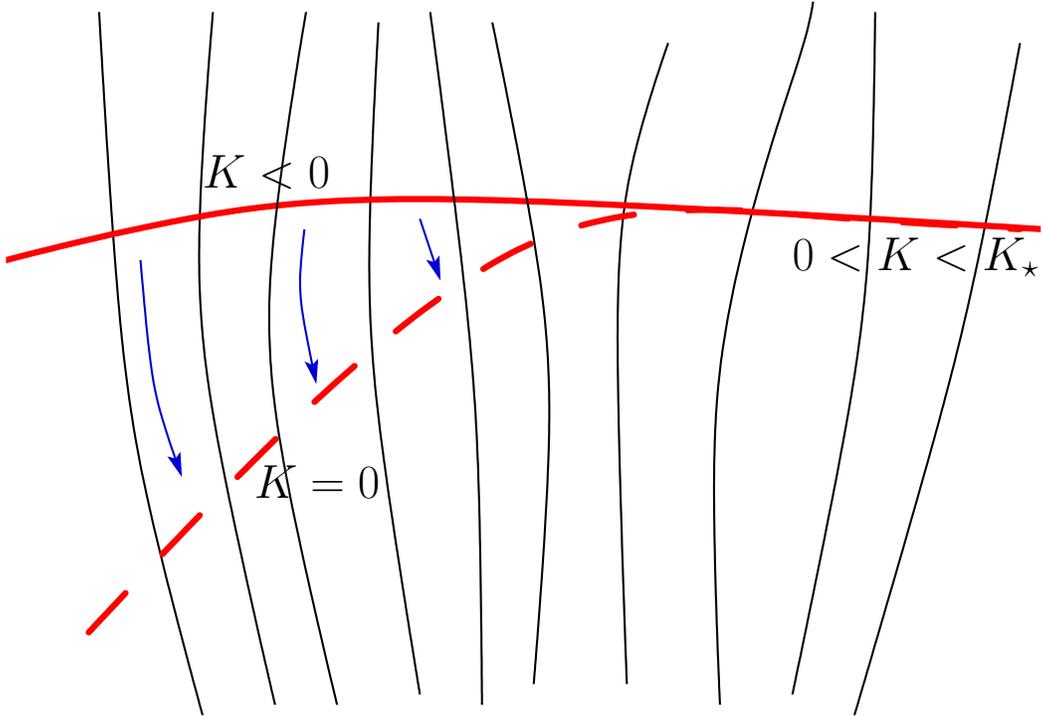}
\end{center}
\caption{
Geometric construction of a surface with $K<K_\star=\sqrt{24\pi G_N T_{\mu \nu} n^\mu n^\nu }$. \label{fig:Kpositive} }
\end{figure}

We have established that all universes of type (iii) and non-empty universes of type (ii) do not have extremal hypersurfaces.  If we now assume the universe is initially expanding ($K>0$ everywhere on some early time slice), we can  show that all slices must have a region that grows at a positive rate $K \geq K_\star$.   Suppose a  hypersurface exists where $K< K_\star$ everywhere.  We can continuously deform the surface in all regions where $K<0$ by pulling it backwards in time along $n^\mu$ until those regions have $K$ arbitrarily close to zero (see Fig.~\ref{fig:Kpositive}). At this point, we have constructed a surface with $|K| < K_\star$ everywhere, violating  (\ref{krho}).

  Now suppose there is a cosmological constant  $\Lambda>0$ (or inflationary potential energy $V$) so that $T_{\mu \nu}n^\mu n^\nu \geq \Lambda$.  Rearranging \eqref{constraint} to
\beq
K^2 = 24 \pi G_N T_{\mu \nu} n^\mu n^\nu + {3 \over 2} \sigma_{\mu \nu} \sigma^{\mu \nu} - {3 \over 2} R^{(3)}
\eeq
and using the fact that universes of type (iii) and (ii) always have a region with $R^{(3)} < 0$ shows that there must be a region with
\beq \label{ds}
K \geq \sqrt{24\pi G_N \Lambda} \equiv K_{\Lambda}.
\eeq
In the flat slicing of de Sitter spacetime $\sigma_{\mu \nu} = R^{(3)} = 0$ and the inequality \eqref{ds} is saturated, so on any  spatial slice there is always a region that expands at least as fast as the flat slicing of de Sitter space does.\footnote{A more formal argument follows from a theorem proven in \cite{gerhardt1983}.
Assuming $K$ grows arbitrarily large near the big bang (which seems inevitable in view of
\eqref{normdeform} since $\sqrt{h} \to 0$), there are early time hypersurfaces with $K > K_{\Lambda}$
everywhere.  Now proceed with a proof by contradiction.  Suppose a  hypersurface with $K < K_{\Lambda}$ everywhere exists at a later time.
Then theorem 6.1 of \cite{gerhardt1983} proves the existence of a  hypersurface in between these two with $K = K_{\Lambda}$
everywhere.  However, in a non-empty universe of type (ii) or (iii), $R^{(3)}<0$ somewhere on this surface, which
violates \eqref{constraint}.  Therefore no hypersurface with $K < K_{\Lambda}$ everywhere can exist.}

\subsection{Black holes}

It is important to note that nothing in our analysis prohibits the
formation of black holes, or any other collapsing or collapsed regions.  What we have shown is only that every spatial slice must contain at least one expanding region with $K > K_{\Lambda} \geq 0$.  In a universe containing both black holes and expanding regions, hypersurfaces that enter the black holes may contain regions of negative $K$, but this does not contradict our theorem so long as $K > K_{\Lambda}$ in at least one region on every time slice.

We demonstrate in  Appendix \ref{BHapp} that the  ``mean curvature flow'' (MCF) procedure
defined above \eqref{volgrowth}   produces surfaces that
avoid the spacelike ``crushing'' singularities \cite{Eardley} of the type that
presumably form inside collapsing black holes.   Intuitively, this is  because these singularities have zero volume and MCF is a  process that locally increases the volume of the hypersurface.

\subsection{Fate of the universe}

It is tempting to  conclude  that the spatial volume of the universe must increase without bound in the future.
 However, it is logically possible that the expanding region shrinks faster than it grows -- that is, we only know that some such neighborhood exists on every slice, but not how large it is, and it could shrink to a point in the asymptotic future.  Were this to be the case it would mean all slicings become singular far enough into the future.  This seems highly implausible, especially in view of the various regularity theorems for mean curvature flow of spacelike hypersurfaces in pseudo-Riemannian manifolds~\cite{ecker1991}.  Furthermore, since we know the total volume cannot reach a maximum at any finite time,  this would be a universe that exists with finite, asymptotically constant volume for eternity.  This is very implausible physically -- such an eternal  finite-volume universe  would be extremely unstable (like the ``loitering" closed universe \cite{1992ApJ...385....1S}).

 Another logical possibility is that the expanding spacetime simply comes to an end on some spacelike surface that is not a zero-volume big crunch.  If the topology is ``flat" or ``open" and the WEC holds, this singular surface would in fact have to have maximal volume (else there would be an earlier non-singular hypersurface with maximum volume, violating the theorem).  It is certainly possible to write a metric for a spacetime with such a singular surface, but certain reasonable extra conditions on the stress-energy  forbid them at least in the homogeneous and isotropic case, and are conjectured to do so more generally~\cite{barrow1986closed}.   These conditions forbid such pathologies as the pressure diverging when the energy density and volume are finite, or the pressure oscillating between finite bounds but with diverging frequency as the surface is approached.
  In any case, such finite-volume singularities seem to us highly artificial and unphysical on general grounds.

 Modulo these concerns, our findings have important consequences for the probability that inflation begins -- and indeed, all initial conditions studied in \cite{East:2015ggf} produced universes with regions that expanded without bound and began inflation.

\subsection{Implications for Inflation}

 We have proven that the weak energy condition, flat or open topology, and initial expansion guarantees that rapid expansion will always continue somewhere (if not everywhere), regardless of any inhomogeneities.  Presumably this leads  to a universe with arbitrarily large volume.  In such a universe vacuum energy must eventually dominate, since all other forms of energy dilute, and therefore inflation will certainly begin if $\Lambda > 0$, or if the initial conditions for the inflaton are such that the inflaton potential potential acts as vacuum energy.

This apparently disposes of the ``initial conditions" issue of requiring large-scale homogeneity in the initial density of the universe~\cite{Vachaspati:1998dy, Ijjas:2015hcc}.  Instead,  the probability for inflation to begin reduces to  estimating the probability for non-spherical topology, along with some measure on the initial conditions for the inflaton.
(It is also worth noting that if the topology is not closed, inflation is actually not required to explain why the universe is so large and so old.  It is however necessary to explain large-scale homogeneity and isotropy.)

Scalar fields (such as the inflaton) violate the weak energy condition if $V(\phi)$ is negative somewhere (of course this is not necessary for inflation, but is nevertheless the case in certain models).  Hence with such potentials our theorem does not strictly apply.  Nevertheless, we expect that it gives some insight into the type of tuning that would be required to avoid inflation in such a model.

Another interesting issue was discovered in \cite{East:2015ggf}  in the case of small-field inflation models.  It is possible to arrange initial conditions such that even though the spatial average of the inflaton is initially on the inflationary plateau, large spatial inhomogeneities the sample the ``steep'' part of the potential have the effect of pulling it rapidly off the plateau and into its minimum before inflation can begin.  The universe expands forever and our theorem is satisfied, but there is little or no inflation.  However this cannot happen in large-field models where the range of the field during inflation $\Delta \phi \simgeq M_{P}$, at least not in the regime that can be described by classical gravity.

\section{Closed universes}

For manifolds of type (i) we cannot prove the universe will not recollapse, but we can use \eqref{constraint} to say something about the energy density relative to the curvature on a putative slice of maximum expansion ($K=0$):
\beq \label{closedconst}
{ 24 \pi G_{N} \Lambda \leq  24 \pi G_{N} T_{\mu\nu}n^\mu n^\nu \leq R^{(3)} \sim 1/l^{2}.}
\eeq
The motivation for the last ``$\sim$''  is a series of mathematical results (see for instance \cite{10.2307/2152760}) showing that Riemannian manifolds with  \emph{Ricci} curvature bounded from below -- all three eigenvalues of the Ricci tensor are greater than some  constant $\lambda > 0$ at every point -- must be compact and have a  diameter $l \leq C \lambda^{-1/2}$, where $C$ is an order one numerical factor and $\lambda$ is the lower bound on the eigenvalues of $R_{i j}$.   Assuming $R^{(3)} \simgeq \lambda$,  if $\Lambda >0$  and at any time the universe grows sufficiently so that its diameter $l  > C (24 \pi G_{N} \Lambda )^{-1/2}$, \eqref{closedconst} cannot be satisfied, so after such a time there cannot be a surface with $K=0$ and the universe will not recollapse.

\section{Non-compact spatial slices}\label{noncompact}

Ordinarily in physics one assumes that boundary conditions (say, periodic) on length scales very much larger than the region of interest cannot affect the physics.  This should hold true  especially in relativistic theories, including general relativity.    Therefore if a  toroidal universe cannot recollapse no matter how large it is, it seems  implausible that a non-compact universe could, at least  on any finite time-scale.

In addition, one  expects non-compact manifolds to have negative  curvature somewhere, because positive curvature reduces the volume (relative to zero curvature).  However, it is in fact possible to find non-compact 3-maniolds with scalar curvature that is positive everywhere -- an example is $S^2 \times R$ with the obvious product metric.  Instead, as mentioned above a sufficient condition is that all eigenvalues of $R_{ij}^{(3)}$ be bounded from below by a positive number.    Hence, for a 3-manifold to be non-compact, at least one eigenvalue of $R_{ij}^{(3)}$ is $\leq0$ (or at least arbitrarily small and asymptoting to zero).  In an inhomogeneous and anisotropic universe with random or highly variable metric and matter configurations, it would seem to require an extreme fine-tuning for the eigenvalues of $R_{ij}^{(3)}$ to be non-positive, while at the same time requiring the sum of the eigenvalues $R^{(3)}$ to be non-negative everywhere in the infinite volume.   This is relevant because if  instead $R^{(3)}<0$ at any point on every slice, our theorem follows and the universe cannot globally recollapse.
In fact with some additional assumptions, this logic can be used to show that a spacetime with a maximal slice must either be compact or violate the weak energy condition (\cite{FrankelGalloway}, Corollary 2).

\section*{Acknowledgements}

The work of MK is supported in part by the NSF through grant PHY-1214302, and he
acknowledges membership at the NYU-ECNU Joint Physics Research Institute in
Shanghai.
LS is supported by DOE Early Career Award DE-FG02-12ER41854.  We would like to thank John Barrow, Jeff Cheeger, Paolo Creminelli, Atish Dabholkar, Will East, Eli Grigsby, Andrei Gruzinov, Andrei Linde, Nemanja Kaloper, Bruce Kleiner, Alberto Nicolis, Massimo Porrati, Neil Turok, Alex Vilenkin, Giovanni Villadoro, Bob Wald, and Matias Zaldarriaga for very useful conversations.

\appendix

\section{Black holes and mean curvature flow}\label{BHapp}

Here, we demonstrate that mean curvature flow (MCF, defined above \eqref{volgrowth})   produces surfaces that
avoid the spacelike ``crushing'' singularities \cite{Eardley} of the type that
presumably form inside collapsing black holes. 

To see this formally, consider starting from a spacelike hypersurface with $K >0$ everywhere.  Evolution by MCF preserves the spacelike nature of the surface because the local volume form is non-decreasing under MCF, but would vanish if the surface became null anywhere.  It also preserves the property that $K>0$ everywhere  (see \emph{e.g.} \cite{gerhardt2006curvature}, Proposition 2.7.1).  Intuitively, this is because because the flow slows to a stop in any region where $K$ approaches zero.  Therefore, starting from a spacelike hypersurface with $K>0$, MCF produces a family of such surfaces.

Naively, one might think that the requirement that $K$ be positive everywhere entirely precludes the surface from entering black hole horizons, but this is not the case.  For instance, the spacelike surface of constant $r = r_{s}-\epsilon$ just inside the event horizon of a Schwarzschild black hole of horizon radius $r_{s}$ is a surface with constant and 
 large positive extrinsic curvature, because the horizon is null and has zero volume.

\begin{figure}
\begin{center}
\includegraphics[width=\columnwidth,draft=false]{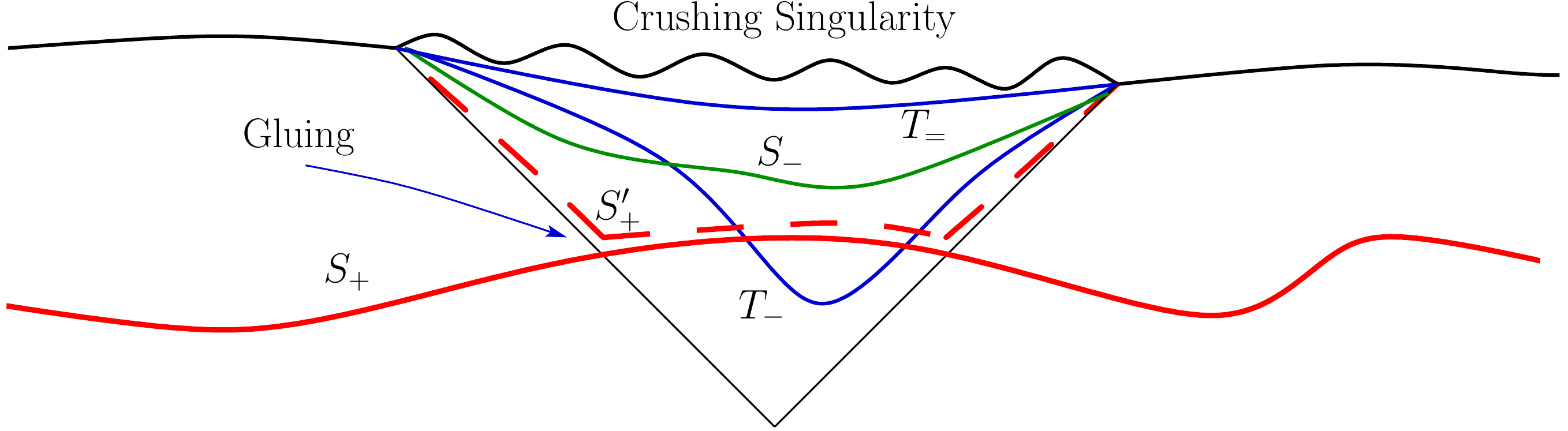}
\end{center}
\caption{
A crushing singularity and the various surfaces used in the argument given in Appendix \ref{BHapp}.  \label{no crush} }
\end{figure}

 Nevertheless, we can use Theorem 2.17 of \cite{Eardley} to prohibit any
$K>0$ surface from approaching arbitrarily close to a crushing singularity.  The argument is a proof by contradiction, as follows.  Assume a hypersurface $S_+$ with $K>0$ exists that comes sufficiently close (to be defined) to the singularity inside a black hole.
To apply the results of \cite{Eardley}, we want to restrict our attention to the interior of the black hole (defined as the interior of the past lightcone of the singularity)
and think of it as a Cauchy complete big crunch cosmology.  Consider a surface $S_+'$ that is entirely contained  inside the hole.  $S_{+}'$ coincides with $S_+$ for the portion of $S_{+}$ that is inside the
horizon.  The rest of $S_{+}'$ consists of the spacelike surface that stays just inside the horizon  (see Fig.~\ref{no crush}).  $S_{+}'$ is a Cauchy surface for the interior of the black hole.  Because $K>0$ everywhere
on $S_+$ and the extrinsic curvature of the surface just inside a null horizon is positive and very large, $K>0$ everywhere on $S_+'$.\footnote{It is straightforward to demonstrate that two  spacelike surfaces such as those in Fig.~\ref{no crush} with $K>0$ can be glued together smoothly in an arbitrarily small region by a surface that is locally a hyperboloid, in such a way that the gluing region also has $K>0$.}

If $S_{+}$ comes sufficiently close to the singularity, it crosses a constant mean curvature (CMC) surface $T_-$ with mean curvature $K_{-}$ satisfying $-\infty < K_{-} < 0$.  The existence of the surface $T_{-}$ at finite distance from the singularity is guaranteed for crushing singularities
by Def.~2.9 of \cite{Eardley}.  Now consider a CMC surface $T_{=}$ with $K_{=} < K_{-} < 0$   that lies entirely to the future of $S_{+}$ -- that is, between $S_{+}$ and the singularity (the
existence of $T_=$ is guaranteed by the same definition).  Theorem 6.1
of \cite{gerhardt1983}  guarantees the existence of a CMC surface $S_-$ with $K=K_{-}$ that lies entirely between $S_+'$ and $T_{=}$.
  But  then $S_{-}$ intersects the region between $T_{-}$ and the singularity, which contradicts Theorem 2.17 of  \cite{Eardley}.  Therefore, $S_+$ cannot come very close to the singularity, and MCF starting from a surface with $K>0$ cannot produce a surface that comes close to a crushing singularity.

One possible caveat in this argument is that the results of \cite{Eardley} assume compactness of the cosmology.  The region behind the horizon of a Schwarzschild black hole is non-compact (because the coordinate $t$ has infinite range).  However, if the cosmology is compact (as we are assuming) and does not grow to infinite volume, the region inside the black hole must be compact as well.  But if the universe does grow to infinite volume, our main result holds true.

\bibliographystyle{JHEP}
\bibliography{ref}

\end{document}